\documentstyle[11pt,aaspp4]{article}
\begin{document}

\title{The DIRECT Project: Influence of Blending on the Cepheid
Distance Scale. I.~Cepheids in M31}

\author{B. J. Mochejska\altaffilmark{1}}
\affil{Copernicus Astronomical Center, 00-716 Warszawa, Bartycka 18}
\affil{\tt e-mail: mochejsk@camk.edu.pl}
\author{L. M. Macri, D. D. Sasselov\altaffilmark{2}, 
K. Z. Stanek\altaffilmark{3}}
\affil{Harvard-Smithsonian Center for Astrophysics, 60 Garden St.,
Cambridge, MA~02138}
\affil{\tt e-mail: lmacri, dsasselov, kstanek@cfa.harvard.edu}
\altaffiltext{1}{Visiting Student, Harvard-Smithsonian Center 
for Astrophysics}
\altaffiltext{2}{Alfred P. Sloan Foundation Fellow}
\altaffiltext{3}{Also N.~Copernicus Astronomical Center,
Bartycka 18, Warszawa PL-00-716, Poland}

\begin{abstract}

We investigate the influence of blending on the Cepheid distance
scale. Blending is the close association of a Cepheid with one or more
intrinsically luminous stars.  High-resolution {\em HST} images are
compared to our ground-based data, obtained as part of the DIRECT
project, for a sample of 22 Cepheids in the M31 galaxy. The average
(median) $V$-band flux contribution from luminous companions which are
not resolved on the ground-based images is about 19\% (12\%) of the
flux of the Cepheid. This is a large effect $-$ at the 10\% level for
distances.  The current Cepheid distance estimates to M31 are all
ground-based, and are thus affected (underestimated). We discuss
indirect methods to find which Cepheids are blended, e.g. by the use
of well-sampled light curves in at least two optical bands.  More
generally, our ground-based resolution in M31 corresponds to the {\em
HST} resolution at about $10\;Mpc$. Blending leads to systematically
low distances in the observed galaxies, and therefore to
systematically high estimates of $H_0$; we discuss the issue and the
implications.

\end{abstract}

\section{Introduction}

As the number of extragalactic Cepheids discovered with {\em HST}
continues to increase and the value of $H_0$ is sought from distances
based on these variables (e.g. Saha et al.~1999; Mould et al.~2000),
it becomes even more important to understand various possible
systematic errors which could affect the extragalactic distance
scale. Currently, the most important systematic is a bias in the
distance to the Large Magellanic Cloud, which provides the zero-point
calibration for the Cepheid distance scale. The LMC distance is very
likely significantly shorter than usually assumed (e.g. Udalski 1998;
Stanek et al.~2000), but it still might be considered uncertain at the
$\sim 10$\% percent level (e.g. Jha et al.~1999).  Another possible
systematic, the metallicity dependence of the Cepheid
Period-Luminosity (PL) relation, is also very much an open issue, with
the empirical determinations ranging from 0 to $-0.4\;mag\; dex^{-1}$
(Freedman \& Madore 1990; Sasselov et al.~1997; Kochanek ~1997;
Kennicutt et al.~1998).

In this paper we investigate a much neglected systematic, that of the
influence of blended stellar images on the derived Cepheid distances.
Although Cepheids are very bright, $M_V\sim -4$ at a period of
$10\;days$, their images when viewed in distant galaxies are likely to
be blended with other nearby, relatively bright stars. We define {\em
blending} as the close projected association of a Cepheid with one or
more intrinsically luminous stars, which can not be detected within
the observed point-spread function (PSF) by the photometric analysis
(e.g., DAOPHOT, DoPHOT). Such blended stars are mostly other young
stars which are physically associated $-$ from actual binary and
multiple systems to companions which are not gravitationally bound to
the Cepheid. Blending is thus a phenomenon different from {\em
crowding} or {\em confusion noise}; the latter occurs in stellar fields
with a crowded and complex background due to the random superposition
of stars of different luminosity. In this paper we are concerned with
blending due to wide unbound systems. Binary Cepheid companions are
well studied (Evans 1992), and do not contribute enough flux to affect
Cepheid distances (Madore 1977), due to the obvious constraints of
coeval stellar evolution. On the other hand, the association of
Cepheids with other luminous stars in wide unbound systems is an
unsolved problem in general.  While such association, i.e. a strong
star-star correlation function, is expected and common to young stars
(Harris \& Zaritsky 1999), the specific case for Cepheids is
unknown. Studies in our Galaxy are very difficult due to the small
sample and existing results, though tantalizing, are inconclusive
(Evans \& Udalski 1994). This could explain the relative neglect of
this issue in recent years, but blending had been of concern for early
studies of Magellanic Cloud Cepheids (DeYoreo \& Karp 1979; Pel, van
Genderen \& Lub 1981), because even faint B-star blends affect the
optical colors of a Cepheid significantly.

We investigate the effects of stellar blending on the Cepheid distance
scale by studying two Local Group spiral galaxies, M31 and M33.  In
this paper we concentrate on M31 (Andromeda Galaxy), located at
approximately $R_{M31}= 780\; kpc$ (e.g. Holland 1998; Stanek \&
Garnavich 1998) from us. As part of the DIRECT project (e.g. Kaluzny
et al.~1998; Stanek et al.~1998) we have collected for this galaxy an
extensive data set, finding among other variables 206 Cepheids.  We
identify some of these Cepheids on archival {\em HST}-WFPC2 images and
compare them to our ground-based data to estimate the impact of
blending on our photometry, taking advantage of their superior
resolution -- the FWHM on the WFPC2 camera corresponds to $\sim
0.4\;pc$ at the distance of M31, compared to $\sim 5\;pc$ for the
ground-based data. The average FWHM on the DIRECT project ground-based
images of M31 is about $1.5\arcsec$, or $\sim 5\;pc$, which
corresponds to the {\em HST}-WFPC2 resolution of $0.1\arcsec$ for a
galaxy at a distance of $10\;Mpc$.  Any luminous star (or several of
them) in a volume of that cross section through the disk (at the
inclination of the galaxy) could be indistinguishable from the Cepheid
and would contribute to its measured flux.  As Cepheids are relatively
young stars, they reside strictly in the midplanes of the disks of
spiral galaxies.

The archival {\em HST}-WFPC2 study of M31 Cepheids from our project
DIRECT was undertaken to improve our distance determination to the
galaxy. The preliminary indirect (via LMC) Cepheid distance we
obtained (Kaluzny et al.  1998; Sasselov et al. 1998) was practically
the same as the Cepheid distance by Freedman \& Madore (1990) of
$R_{M31}= 770~\pm 25\; kpc$.  Our findings of blending now indicate
that these distance estimates should be corrected upward.

We describe the ground-based and {\em HST} data and the applied
reduction procedures in Section 2.  In Section 3 we discuss the task
of identifying Cepheids on {\em HST} WFPC2 images. In Section 4 we
present the blending catalog of Cepheids and discuss it in Section
5. In Section 6 we describe the effect of blending on the light and
color curves of Cepheids and how these curves can be used to detect
blends. In Section 7 we discuss the effectiveness of some of the blend
analysis methods encountered in the literature.  The concluding
remarks are to be found in Section 8.

\section{Observations and Data Reduction}

\subsection{Ground-based Data}

The ground-based data were obtained as part of the DIRECT project
between September 1996 and October 1997 during 95 full/partial nights
on the F. L. Whipple Observatory 1.2~m telescope and 36 full nights on
the Michigan-Dartmouth-MIT 1.3~m telescope. Six $11'\times11'$ fields
with a scale of 0.32 \arcsec/pixel were monitored: four of them (A--D)
concentrated on the rich spiral arm in the northeast part of M31, one
(E) close to the bulge of M31 and one (F) containing the giant star
formation region known as NGC 206. Fields A--D and F have been reduced
and the $BVI$ photometry of Cepheid variables published in Stanek et
al.~1998 (hereafter, Paper II), Kaluzny et al.~1998 (Paper I), Stanek
et al.~1999 (Paper III), Kaluzny et al.~1999 (Paper IV) and Mochejska
et al.~1999 (Paper V), respectively. The applied reduction,
calibration and variable selection procedures are discussed therein,
particularly in Paper I, where full details are provided. A total of
206 Cepheids were found: 43 in field A, 38 in B, 35 in C, 38 in D and
52 in field F.

\subsection{HST data}

The archival {\em HST}-WFPC2 data used in this paper were retrieved
from the Hubble Data Archive. We selected images overlapping our M31
ground-based data, taken in filters F336W (roughly $U$), F439W, F450W
($\sim B$), F555W, F606W ($\sim V$) and F814W ($\sim I$). The pixel
scales of the Wide Field (WF) and Planetary Camera (PC) chips are
0.0996 and 0.0455 \arcsec/pixel, respectively. The full list of
exposures is provided in Table \ref{tab:dat}, along with the proposal
ID, dataset name, equatorial coordinates of the frame centers, filter
and exposure time information.

\begin{small}
\tablenum{1}
\begin{planotable}{llcccc}
\tablewidth{35pc}
\tablecaption{\sc The full list of {\em HST} images containing DIRECT Cepheids}
\tablehead{\colhead{Proposal ID} & \colhead{Dataset Name} &
\colhead{$\alpha_{J2000.0}$} & \colhead{$\delta_{J2000.0}$} &
\colhead{Filter} & \colhead{Exp.~Time (s)} }
\startdata
5911 & U2Y20103T & 00 44 44.17 & 41 27 33.88 & F336W & 400 \\
5911 & U2Y20105T & 00 44 44.23 & 41 27 33.86 & F439W & 160 \\
5911 & U2Y20106T & 00 44 44.23 & 41 27 33.86 & F555W & 140 \\
5911 & U2Y20203T & 00 44 49.28 & 41 28 59.04 & F336W & 400 \\
5911 & U2Y20205T & 00 44 49.34 & 41 28 59.03 & F439W & 160 \\
5911 & U2Y20206T & 00 44 49.34 & 41 28 59.03 & F555W & 140 \\
5911 & U2Y20303T & 00 44 57.57 & 41 30 51.68 & F336W & 400 \\
5911 & U2Y20305T & 00 44 57.63 & 41 30 51.65 & F439W & 160 \\
5911 & U2Y20306T & 00 44 57.63 & 41 30 51.65 & F555W & 140 \\
5911 & U2Y20403T & 00 45 09.20 & 41 34 30.56 & F336W & 400 \\
5911 & U2Y20405T & 00 45 09.25 & 41 34 30.72 & F439W & 160 \\
5911 & U2Y20406T & 00 45 09.25 & 41 34 30.72 & F555W & 140 \\
5911 & U2Y20503T & 00 45 11.89 & 41 36 56.86 & F336W & 400 \\
5911 & U2Y20505T & 00 45 11.95 & 41 36 57.02 & F439W & 160 \\
5911 & U2Y20506T & 00 45 11.95 & 41 36 57.02 & F555W & 140 \\
6038 & U2YE0603T & 00 44 51.22 & 41 30 03.72 & F555W & 160 \\
6038 & U2YE0605T & 00 44 51.22 & 41 30 03.72 & F439W & 600 \\
6038 & U2YE060BT & 00 44 51.22 & 41 30 03.72 & F336W & 900 \\
6038 & U2YE060DT & 00 44 51.22 & 41 30 03.72 & F336W & 900 \\
5237 & U2AB0101T & 00 40 29.40 & 40 43 58.28 & F555W & 200 \\
5237 & U2AB0102T & 00 40 29.40 & 40 43 58.28 & F555W & 200 \\
5237 & U2AB0103T & 00 40 29.40 & 40 43 58.28 & F814W & 200 \\
5237 & U2AB0104T & 00 40 29.40 & 40 43 58.28 & F814W & 200 \\
5237 & U2AB0105T & 00 40 29.40 & 40 43 58.28 & F336W & 600 \\
5237 & U2AB0106T & 00 40 29.40 & 40 43 58.28 & F336W & 600 \\
5494 & U2G20701T & 00 40 33.17 & 40 45 38.97 & F606W & 350 \\
8059 & U4WOAH02R & 00 40 10.11 & 40 46 08.91 & F606W & 400 \\
8059 & U4WOAH04R & 00 40 10.11 & 40 46 08.91 & F450W & 600 \\
8059 & U4WOAH05R & 00 40 10.11 & 40 46 08.91 & F814W & 100 \\
8059 & U4WOAH06R & 00 40 10.11 & 40 46 08.91 & F814W & 300 \\
8061 & U4X1OF01R & 00 40 10.11 & 40 46 08.91 & F606W & 400 \\
\enddata
\label{tab:dat}
\end{planotable}
\end{small}

The data we obtained had already passed through the standard
preliminary processing and calibration procedures prior to its
placement in the Archive. The standard pipeline calibration is fully
described in the {\em HST} Data Handbook.

The first two steps in our reduction procedure were to mark the bad
pixels on the images and to compensate for the fact that pixels on the
edges and corners of the CCD receive fewer photons due to the
geometric distortion in the WFPC2 optics. For each image a mask was
created from the data quality file retrieved from the Archive and a
vignetting mask generated by Stetson (1998) and then used to mark bad
pixels and vignetted regions. To restore the integrity of the flux
measurements the images were multiplied by a pixel-area map,
originally created by Holtzman et al.~(1995) and renormalized to the
median pixel area on each chip by Stetson (1998). These tasks were
accomplished under IRAF\footnote{IRAF is distributed by the National
Optical Astronomy Observatories, which are operated by the Association
of Universities for Research in Astronomy, Inc., under cooperative
agreement with the NSF.}.

In the next step of the reduction procedure pairs (or multiplets) of
images were selected, taken in the same filters, having identical
center coordinates as well as similar exposure times.  The {\bf crrej}
task under IRAF was used to combine the images and remove cosmic
rays. When multiple images for a given field were not available,
single images were used for photometry.

The photometry was extracted by means of the DAOPHOT/ALLSTAR package
(Stetson 1987, 1992). Stars were identified using the FIND subroutine
and aperture photometry was done on them with the PHOT subroutine.  We
used the point-spread functions (PSF) derived individually for each
filter and chip of the WFPC2 camera, kindly provided to us by
P.~B.~Stetson (private communication) to obtain the ALLSTAR profile
photometry. After the initial ALLSTAR run, FIND was ran again on the
star subtracted images, to identify stars that were missed on the
first pass. Aperture photometry was obtained for them with PHOT,
followed by profile photometry with ALLSTAR. The two star lists were
merged and used as input to ALLSTAR to obtain the final photometry.

It should be noted that the {\em HST} photometry has not been
calibrated to any standard system and therefore instrumental
magnitudes are used throughout this paper. This has, however, no
bearing on the results presented in this paper, since they are
strictly based on differential photometry.

\section{The Identification of DIRECT Cepheids in {\em HST} Data}

The preliminary identification of DIRECT Cepheids on the {\em HST} frames
was accomplished by a visual comparison of the {\em HST} data matched via
World Coordinate System (WCS) information to our ground-based $V$ template
image for each field, using SAOimage ds9\footnote{SAOimage ds9 was
developed under a grant from NASA's Applied Information System Research
Program (NAG5-3996), with support from the Chandra Science Center
(NAS8-39073)}. The WCS information for the ground-based template image
header was obtained with a program written by Mink (1997; 1999) by matching
stars from the image to the USNO A2.0 Catalog stars (Monet et al.~1996). We
have matched a total of 22 Cepheids to {\em HST} data: six in our field B,
four in C (two of them on two different overlapping fields) and 12 in field
F (one found on two different fields). This constitutes $\sim 18\%$ of the
125 Cepheids in fields B, C and F.

\begin{figure}[!t]
\plotfiddle{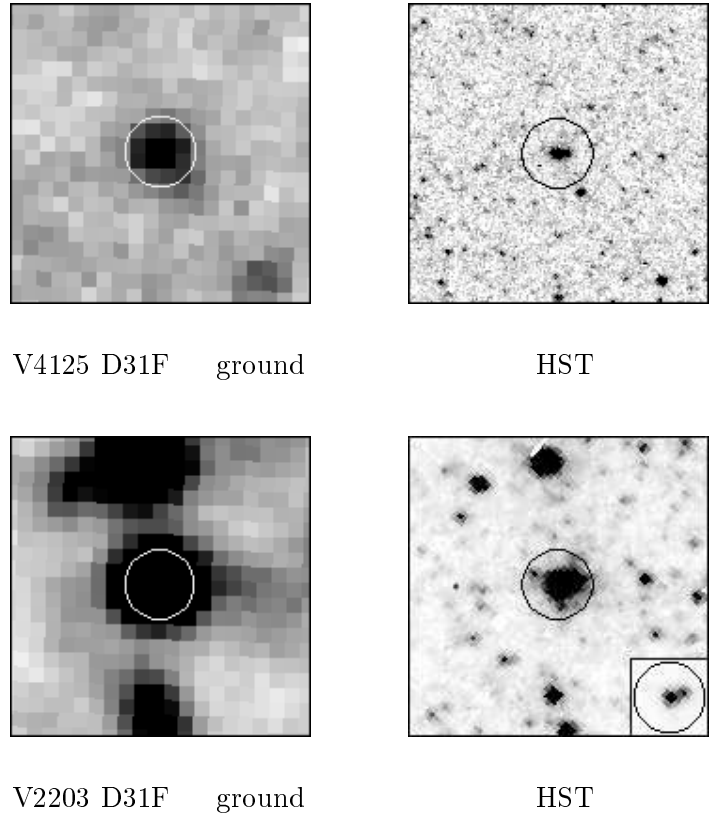}{10.5cm}{0}{100}{100}{-205}{-370}
\caption{A comparison of ground-based and {\em HST} data for two
Cepheids: V4125 D31F and V2203 D31F. The images plotted in the left
panels are taken from the $V$ template. The circles, centered on the
Cepheids, have a radius of $0.75\arcsec$. The inset in the lower right
panel has different IRAF z-scale limits to show the two blended stars
separately.}
\label{fig:cfig}
\end{figure}

In some cases, when a Cepheid which appeared to be a single star on
the $V$ template was resolved into multiple stars on an {\em HST}
image, it was difficult to distinguish which star was the Cepheid on
the basis of the template image alone. Two such cases are shown in
Figure \ref{fig:cfig}. The images plotted in the left panels are taken
from the $V$ template, created by averaging together two images with
the best seeing (FWHM $\sim 1\arcsec$).  The image in the upper right
panel was taken with the PC chip of the WFPC2 camera while the one in
the lower right panel comes from a WF chip.  The inset has different
IRAF z-scale limits to show the two blended stars separately. The
circles, centered on the Cepheids, have a radius of $0.75\arcsec$. The
Cepheid V4125 D31F is shown in the upper panels of the Figure
\ref{fig:cfig}. On the ground-based template it appears as a single
star, while on the PC chip it is resolved into two objects, one 3
magnitudes dimmer than the other. A third star appears below and to
the right, just outside the circle, fainter by almost 2 magnitudes
from the brighter component of the doublet. That star was not
identified on the ground-based template image, but after subtracting
the Cepheid a brighter spot is visible at that location. The lower
panels show the case of V2203 D31F, where the Cepheid, appearing as a
single star on the ground $V$ template image, is resolved by {\em HST}
into two stars, differing by one magnitude in brightness.

\begin{figure}[t]
\plotfiddle{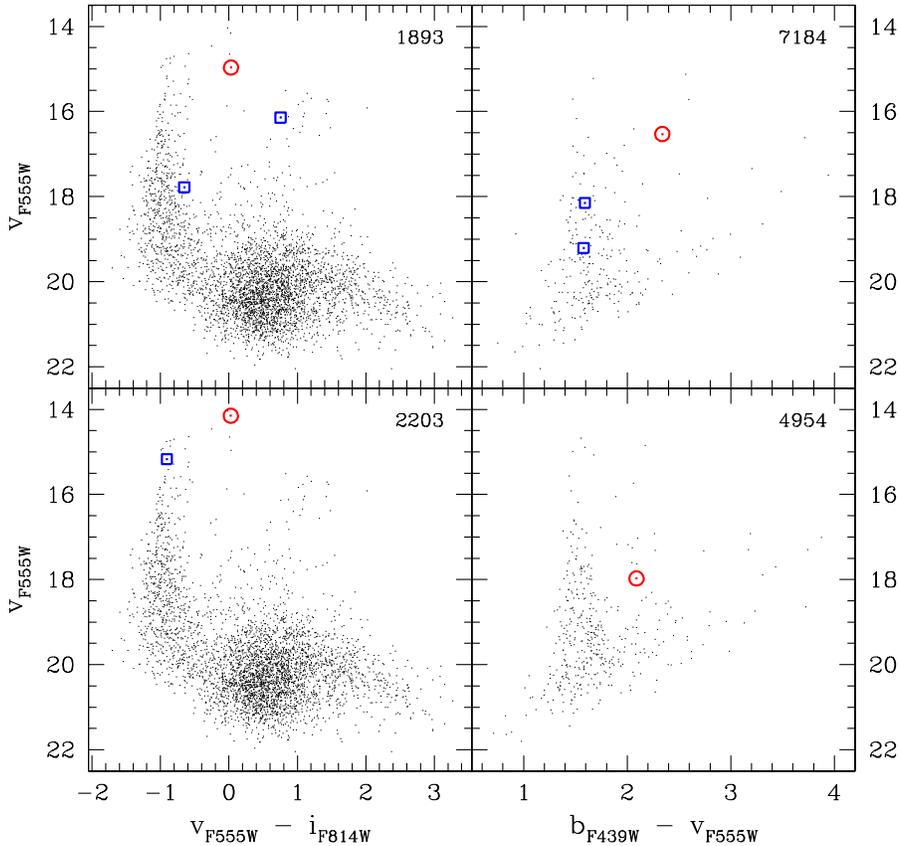}{10cm}{0}{60}{60}{-195}{-97}
\caption{Selected color-magnitude diagrams for Cepheids and their
companions within $0.75\arcsec$ based on {\em HST} data. The Cepheids
are denoted by circles and their companions by squares. Stars from the
same chip are plotted in the background.}
\label{fig:cmd}
\end{figure}

To help confirm the Cepheid nature of the selected objects,
instrumental color-magnitude diagrams (CMDs) were constructed from
{\em HST} data, whenever photometry in two bands was available. A few
representative CMDs are shown in Figure~\ref{fig:cmd}, ($v_{F555W},
v_{F555W}-i_{F814W}$) in the left panels and ($v_{F555W},b_{F439W}
-v_{F555W}$) in the right panels. The Cepheids are denoted by circles
and their companions by squares. Stars from the same chip are plotted
in the background for reference. The upper left panel shows a case
where there is a substantial contribution of flux from a red giant
companion ($(\frac{f}{f_C})_V=34\%$, $(\frac{f}{f_C})_I=67\%$) which
is seen at a distance of $0.2\arcsec$ from the Cepheid V1893 D31F. The
other companion is a blue star at a distance of $0.4\arcsec$ which
appears to be located at the red edge of the main sequence (also in
the $v_{F555W},u_{F336W}-v_{F555W}$ CMD which is not shown). The lower
left panel presents a case where the Cepheid V2203 D31F has a luminous
blue main sequence companion with $(\frac{f}{f_C})_V=40\%$ and
$(\frac{f}{f_C})_I=17\%$ which are separated by $0.3\arcsec$. The
upper right panel shows the Cepheid V7184 D31B with two blue main
sequence companions, the more luminous one, located $0.4\arcsec$ away,
contributing $45\%$ as much light as the Cepheid in the $B$ band and
$23\%$ in $V$. A typical situation where none of the companions (in
this case one) had $b$-band photometry (see Table \ref{tab:cep}) is
illustrated in the lower right panel of Figure~\ref{fig:cmd}, where
the Cepheid V4954 D31B is shown.

\section{M31 Blending Catalog}
\subsection{The Catalog}

The criterion for blending was determined empirically, by careful
examination of ground-based and {\em HST} data for the same
Cepheids. We consider a star to be blended with the Cepheid if it is
located at a distance of less than $0.75\arcsec$ from it and is not
detected by DAOPHOT in our ground-based images. The choice of maximum
distance was motivated by the typical full width at half maximum
(FWHM) in our ground-based images ($\sim 1.5\arcsec$).

Due to the relatively small number of Cepheids having {\em HST} data,
it was deemed worthwhile to examine them on the {\em HST} images in
detail to see whether all of the companions had been identified by
DAOPHOT and to check for the possibility of false detections (cosmic
rays in case of single images, bad columns, etc.). In the latter case
the object was removed from the list. Not much could be done in the
former case, although that proved not to be a problem and only in a
few of the 55 examined images there were very faint companions which
eluded detection. It must be noted that the FWHM of the PSF on the WF
chip is of the order of 1 pixel and so the probability of the
detection of a faint star may depend on the way its light is
distributed over the pixels.  One exception, however, is V1893 D31F,
which DAOPHOT failed to recognize as a star on one of the single F606W
filter images, along with its companion 2 pixels ($0.2\arcsec$)
away. No problem was encountered on the other three combined images.

We estimate that our data is fairly complete for companions
contributing at least $4\%$ of the flux of the Cepheid in the V band
(filters F555W and F606W). We have used this somewhat arbitrary cutoff
in evaluating the sum $S_F$ of all flux contributions in filter $F$
normalized to the flux of the Cepheid:
\begin{equation}
S_F= \sum_{i=1}^{N_F}\frac{f_i}{f_C}
\label{eq:sv}
\end{equation}
where $f_i$ is the flux of the i-th companion, $f_C$ the flux of the
Cepheid on the {\em HST} image and $N_F$ the total number of
companions. In Table \ref{tab:cep} we present the results for each of
the 22 Cepheids found in the {\em HST} images: the name, the mean $V$,
$I$ and $B$ magnitudes taken from Papers II, III and V, the number of
companions $N_F$ and their total flux contribution $S_F$ in the $V$,
$I$, $B$ and $U$ bands respectively. Unless noted otherwise, the $V$,
as in $N_V$ and $S_V$, refers to filter F555W, $I$ to F814W, $B$ to
F439W and $U$ to F336W. For the two Cepheids identified on two {\em
HST} images, the average values of $S_F$ are listed.

\begin{small}
\tablenum{2}
\begin{planotable}{llllllllllll}
\tablewidth{35pc}
\tablecaption{\sc The Cepheid blending catalog}

\tablehead{\colhead{Name} & \colhead{$\langle V \rangle$} &
\colhead{$\langle I \rangle$} & \colhead{$\langle B \rangle$} &
\colhead{$N_V$} & \colhead{$S_V$} & \colhead{$N_I$} & 
\colhead{$S_I$} & \colhead{$N_B$} & \colhead{$S_B$} &
\colhead{$N_U$} & \colhead{$S_U$}}

\startdata
V7184  D31B& 19.15& 18.45&\nodata& 3& 0.40&  &\nodata& 2& 0.62  & 3& 2.55  \\
V6379  D31B& 20.65& 19.39&\nodata& 3& 0.27&  &\nodata& 0& 0.00  &  &\nodata\\
V7209  D31B& 20.07& 19.14&\nodata& 1& 0.12&  &\nodata& 0& 0.00  & 1& 0.96  \\
V5646  D31B& 20.90& 19.51&\nodata& 1& 0.05&  &\nodata& 0& 0.00  &  &\nodata\\
V4954  D31B& 20.86& 19.98&\nodata& 1& 0.10&  &\nodata& 0& 0.00  & 0& 0.00  \\
V6872  D31B& 21.60& 20.62&\nodata& 1& 0.05&  &\nodata& 0& 0.00  & 0& 0.00  \\
V14487 D31C& 19.95& 18.95& 20.75 & 2& 0.12&  &\nodata& 0& 0.00  & 1& 0.60  \\
V13705 D31C& 21.50& 19.98& 22.83 & 0& 0.00&  &\nodata&  &\nodata&  &\nodata\\
V14661 D31C& 21.39& 19.70& 22.76 & 2& 0.30&  &\nodata& 0& 0.00  &  &\nodata\\
V14361 D31C& 21.21& 19.32& 22.36 & 3& 0.27&  &\nodata& 0& 0.00  &  &\nodata\\
V4125  D31F& 20.46& 19.63& 21.11 &1&0.06&   3& 0.25  &  &\nodata& 0& 0.00  \\
V3289  D31F& 20.89& 19.76& 20.77 &2&0.11&   3& 0.27  &  &\nodata& 1& 0.53  \\
V3550  D31F& 20.55& 19.85& 21.03 &2&0.47&   0& 0.00  &  &\nodata& 1& 6.33  \\
V1893  D31F& 18.79& 17.47& 19.54 &2&0.41&   1& 0.66  &  &\nodata& 1& 0.33  \\
V2203  D31F& 17.94& 17.17& 18.32 &1&0.39&   1& 0.17  &  &\nodata& 3& 2.88  \\
V3860  D31F& 21.24& 19.65& 22.08 &0&0.00&   3& 0.63  &  &\nodata& 0& 0.00  \\
V3441  D31F& 21.28& 20.52& 21.18 &2&0.26$^1$&&\nodata&  &\nodata&  &\nodata\\
V2320  D31F& 20.27& 19.54& 20.60 &0&0.00$^1$&&\nodata&  &\nodata&  &\nodata\\
V1633  D31F& 21.30& 19.85&\nodata&1&0.10$^1$&&\nodata&  &\nodata&  &\nodata\\
V1599  D31F& 20.94& 19.86& 21.78 &4&0.34$^1$&&\nodata&  &\nodata&  &\nodata\\
V1549  D31F& 20.72& 19.68& 21.49 &4&0.28$^1$&&\nodata&  &\nodata&  &\nodata\\
V7074  D31F& 20.78& 19.70& 21.51 &0&0.00$^1$& 0& 0.00 &0&0.00$^2$&  &\nodata\\

\enddata
\tablecomments{$^1$ results obtained in the F606W filter\\
\hspace*{45pt} $^2$ results obtained in the F450W filter}
\label{tab:cep}
\end{planotable}
\end{small}

\begin{figure}[t]
\plotfiddle{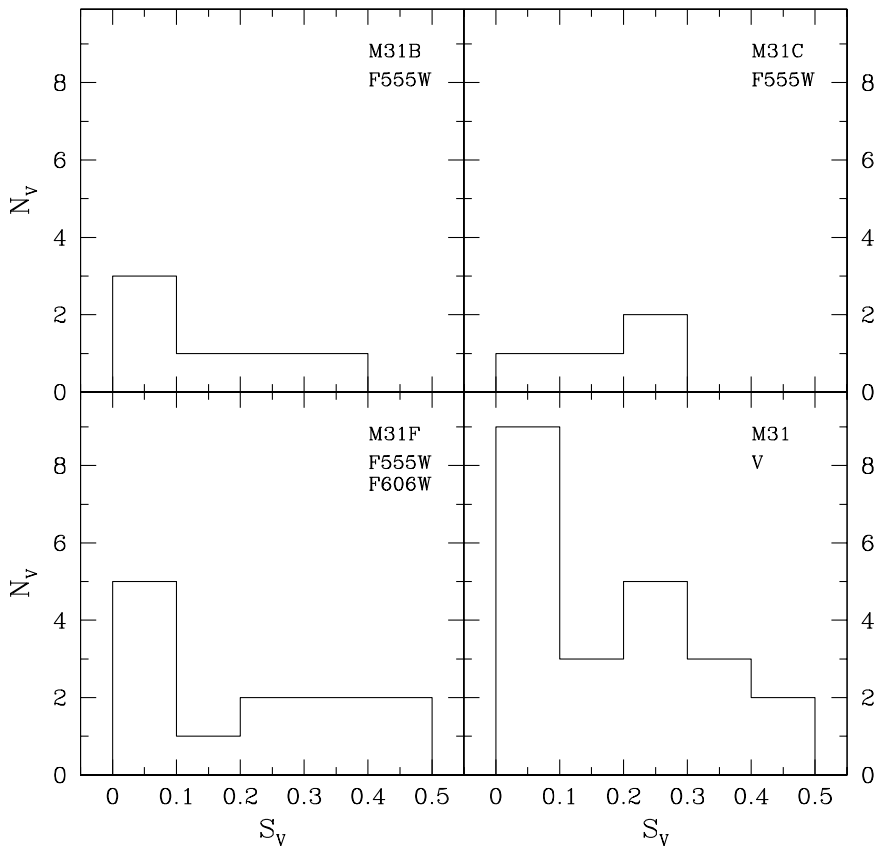}{10cm}{0}{60}{60}{-195}{-97}
\caption{A histogram showing the number of Cepheids $N_V$ as a
function of $S_V= \sum_{i=1}^{N_F}\frac{f_i}{f_C}$, the sum of flux
contributions from its companions in the $V$ band (filters F555W and
F606W), normalized to the flux of the Cepheid.  The upper left and
right panels and the lower left panel are for fields
B, C and F, respectively; the lower right panel is the combined data.}
\label{fig:hist}
\end{figure}

The catalog of Cepheid blending in M31 is illustrated in the following
two figures.
Figure~\ref{fig:hist} shows a histogram of the number of Cepheids
$N_V$ as a function of $S_V$ (Eq.~\ref{eq:sv}), the sum of flux
contributions from its companions in the $V$ band (filters F555W and
F606W), normalized to the flux of the Cepheid. The upper left and
right panels and the lower left panel show the histograms for fields
B, C and F, respectively. In the lower right panel the combined data
is shown. The width of each bin is 0.1 and the first one starts at 0.
For further discussion on Figure~\ref{fig:hist} see \S 5.

\begin{figure}[t]
\plotfiddle{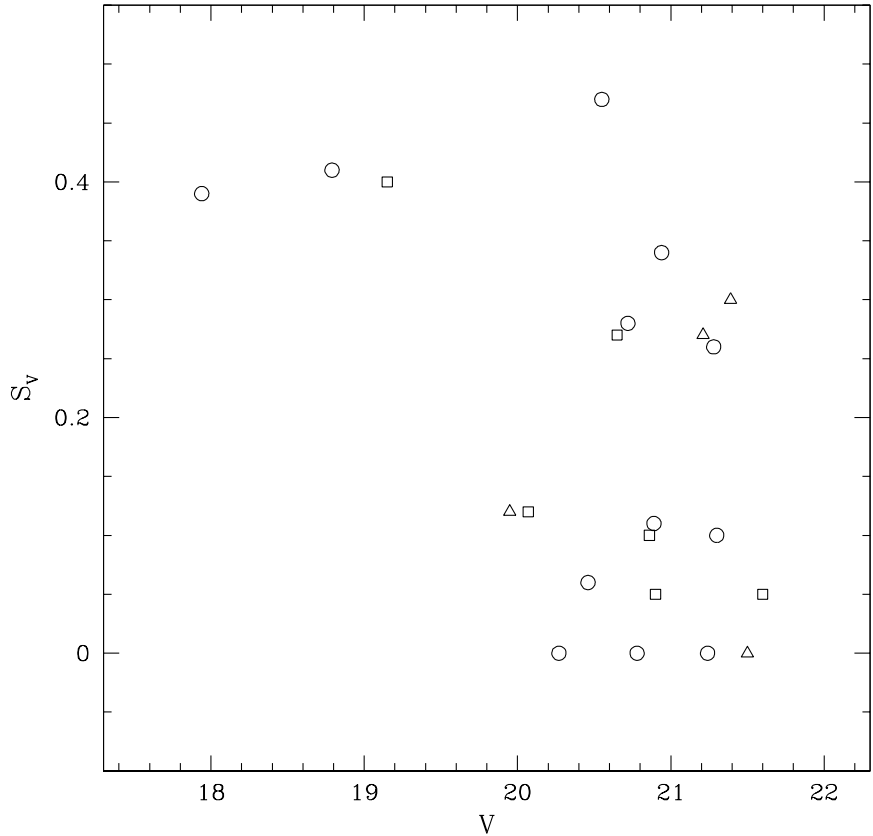}{10cm}{0}{60}{60}{-195}{-97}
\caption{Cepheid blending is not a function of magnitude, $i.e.$,
luminosity, at the same distance in M31.
A diagram of the flux contribution from companions $S_V=
\sum_{i=1}^{N_F}\frac{f_i}{f_C}$ as a function of the $V$ magnitude of
the Cepheid obtained from ground-based data for all 22 DIRECT Cepheids
present in {\em HST} data.  Field B Cepheids are denoted by squares,
field C by triangles and field F by circles.}
\label{fig:vfv}
\end{figure}

A diagram showing the flux contribution from companions $S_V$ as a
function of the $V$ magnitude of the Cepheid obtained from
ground-based data (Papers II, III and V) is presented in Figure
\ref{fig:vfv}. Field B Cepheids are denoted by squares, field C by
triangles and field F by circles.

\subsection {The Blended Cepheids and their Environments}

\begin{figure}[t]
\plotfiddle{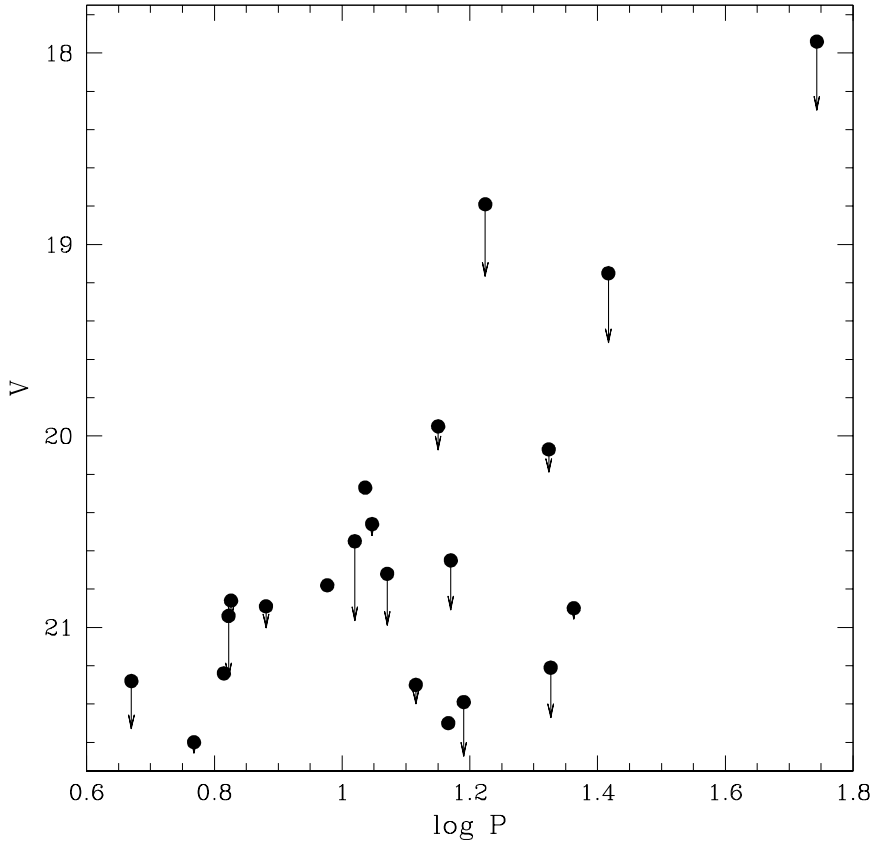}{10cm}{0}{60}{60}{-195}{-97}
\caption{The $\log P$ vs. $V$ diagram for the 22 Cepheids found in
{\em HST} data (reddening and the random-phase nature of the {\em HST} data
are not accounted for). The arrows illustrate the shift in $V$ when the effects
of blending are taken into account. The effect is significant, but it is
not a function of period.}
\label{fig:pl}
\end{figure}

The first step in using the derived blending is to see what is
the effect on the Cepheid period-luminosity (P-L) relation.
Figure \ref{fig:pl} illustrates the influence of blending on the
location of the Cepheids on the P-L diagram. The solid circles show
the original locations of the Cepheids, based on their mean $V$
magnitudes obtained from the ground-based data. The arrows illustrate
the shift in $V$ when the effects of blending are taken into account.
It has to be kept in mind that neither the differential reddening in
M31 nor the random-phase nature of the {\em HST} data have been
accounted for, thus there is significant scatter in the diagram.
However, the direction and the magnitude of the effect is obvious.

\begin{figure}[t]
\plotfiddle{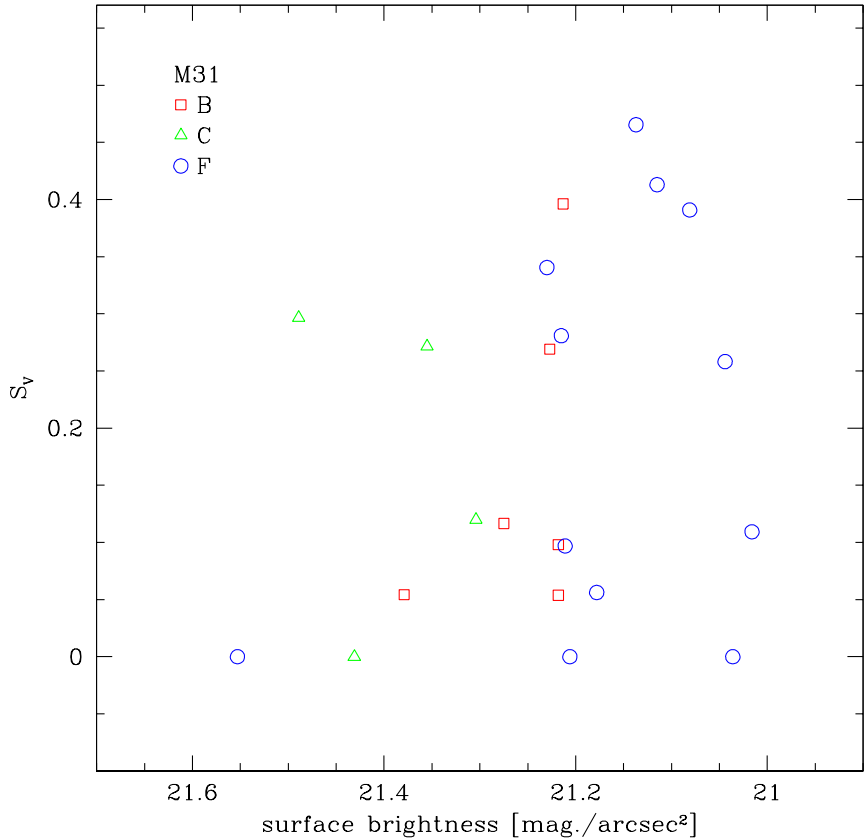}{10cm}{0}{60}{60}{-195}{-97}

\caption{Blending $S_V$ as a function of the surface brightness around
the M31 Cepheids on the ground-based templates. Field B Cepheids are
denoted by squares, field C by triangles and field F by circles. Cepheid
blending does not seem to correlate with surface brightness.}

\label{fig:sky_sv}
\end{figure}

A pattern in the spatial distribution of Cepheid blending in M31
could show us a correlation between blending and crowding. We find no such
correlation -- Cepheids in environments of different surface brightness
show roughly the same frequency and amount of blending, thus distinguishing
the phenomenon of blending from crowding.
In Figure \ref{fig:sky_sv} the blending parameter $S_V$ is plotted as
a function of the surface brightness around the Cepheid for which it
was determined. The surface brightness was taken to
be the mode within a 20 pixel radius on the DIRECT template frames. We
used a bin width of 1 ADU for the histogram to compute the value of the
mode, smoothed with a flat-topped rectangular kernel (boxcar) filter,
11 units in length.  After correcting for the sky level, the
instrumental surface brightness values were converted to
mag/\sq \arcsec using 12-17 fairly bright isolated reference
stars with known standard magnitudes by means of the following
formula:
\[ SB\;[{\rm mag}/\sq \arcsec] = m_{ref} + 2.5 \log 
(s^2 I_{ref} / I_{SB})\] 
where $I_{ref}$ and $m_{ref}$ are the flux in ADU/\sq\ pixel
and the corresponding magnitude of the reference star, $I_{SB}$ is the
surface brightness expressed in ADU/\sq\ pixel and $s$ is the
pixel scale of the chip. The $rms$ scatter around the value of the
average ranged from 0.025 to 0.061 mag/\sq\arcsec.

\begin{figure}[t]
\plotfiddle{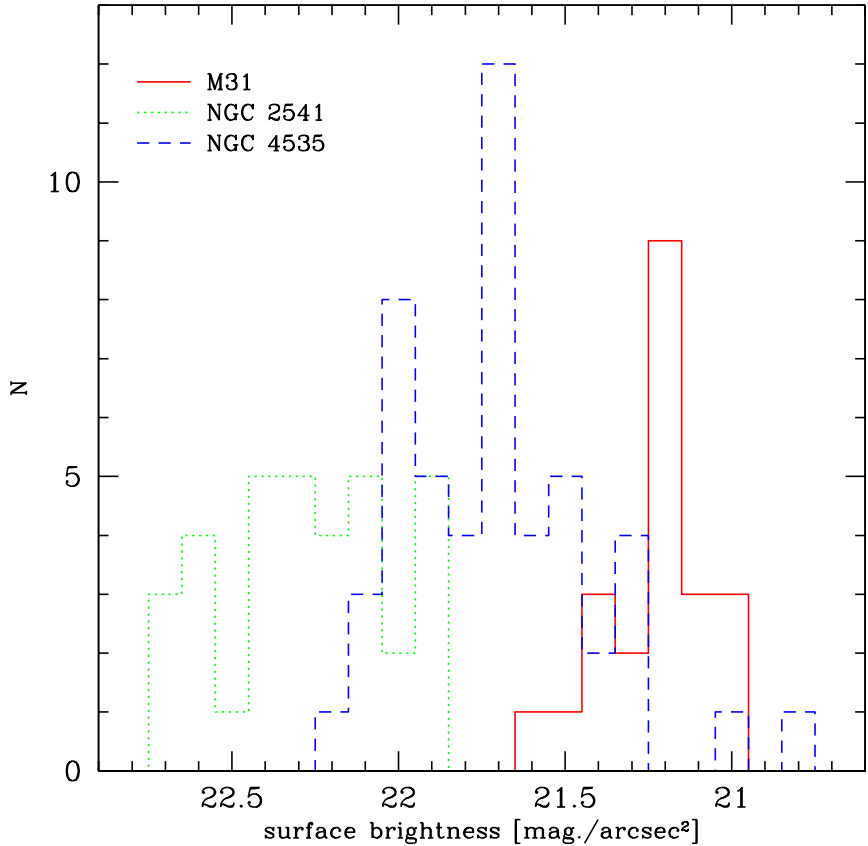}{10cm}{0}{60}{60}{-195}{-97}

\caption{The surface brightness around Cepheids in M31, NGC 4535 and
NGC 2541. The Cepheids that we observe in M31 do not reside in 
anomalously high surface brightness areas.} 

\label{fig:sky_sv_cmp}
\end{figure}

In order to give these numbers some perspective, we have computed surface
brightness around Cepheids in two galaxies which straddle their range, NGC
2541 and NGC 4535 (Fig. \ref{fig:sky_sv_cmp}).  We find that the Cepheids
we discover in M31 reside in environments of surface brightness typical of
spiral galaxies, despite the high inclination of M31. The {\em HST} data
for these galaxies, observed as part of the HST Key Project on the
Extragalactic Distance Scale (Ferrarese et al. 1998, Macri et al. 1999),
were obtained from the Hubble Data Archive. Since we were unable to
estimate the sky level for those images, we tried to choose epochs where
the sky level would be the lowest. We chose the 1995 Nov 20 epoch for NGC
3541 and the 1996 May 24 epoch for NGC 4535. The surface brightness values
were computed using the same method as for our M31 Cepheids and were
re-expressed in units of mag/\sq\arcsec using the zeropoints
provided in the HST Data Handbook.

\section{Discussion of Blending Properties}

Before attempting to draw any far-reaching conclusions from
Figs.~\ref{fig:hist}-\ref{fig:sky_sv_cmp} or the catalog itself
(Tab.~\ref{tab:cep}), we must stress that they are based on a fairly
small sample of Cepheids and therefore subject to statistical
uncertainties.  An illustration of this effect is the small number of
counts in the second bin of the combined histogram
(Fig.~\ref{fig:hist}), compared to the third and, especially, the
first bin, which could be an artifact of small number statistics.  An
encouraging fact is that the distributions in separate fields,
especially B and F, appear to be similar to a substantial degree in
their overall shape and spanned range of $S_V$. In the combined
histogram there is a decreasing trend in the number of cases $N_V$
with increasing flux contribution from blending, $S_V$.

The diagram in Fig.~\ref{fig:vfv} conveys some of the information
contained in the previous figure -- a lack of cases with $0.12 \leq
S_V \leq 0.23$ is readily apparent, which, as mentioned above, is
believed to result from the fairly small size of the sample. If we
attribute the gap to statistical effects, then Cepheids fainter than
$V\sim20$ appear to populate a wide range of $S_V$, from 0 up to
approximately 0.35, with one object at $S_V\sim0.47$.  Furthermore, it
is observed that in none of the fields do Cepheids show a tendency to
favor any particular value of $S_V$, as also apparent from
Figure~\ref{fig:hist}.

Another striking feature of the diagram is that all of the three
Cepheids brighter than $V\sim20$ have $S_V\sim0.4$, although it has to
be emphasized that this sample is much too small to draw any definite
conclusions. One possible explanation may be ventured, however, which
may hold true in these particular cases, but does not have to be the
rule for the whole population of bright Cepheids.  The most luminous
Cepheids are the most massive ones and, hence, the youngest. The group
or association of stars in which they had formed will have had less
time to disperse. This may manifest itself in a somewhat higher
spatial density of stars and thus would increase the probability of
blending.

Figure \ref{fig:sky_sv} shows the blending parameter $S_V$ of the M31
Cepheids as a function of the local surface brightness.  As there is a
gap at $S_V\sim0.2$, the separation into blended ($S_V > 0.25$) and
unblended Cepheids ($S_V < 0.15$) is readily discernible in the
diagram.  No correlation between the blending parameter $S_V$ and the
underlying surface brightness is apparent.

To put the surface brightness values around the M31 Cepheids into
perspective, we plot the surface brightness values computed for the
NGC 4535 and NGC 2541 Cepheids (Fig. \ref{fig:sky_sv_cmp}). The range
of surface brightness of NGC 2541 falls below the range of our M31
Cepheids. There is a 26\% overlap between the NGC 4535 and M31 data --
13 out of the 50 Cepheids in NGC 4535 are located in regions with
surface brightness in the range covered by our M31 Cepheids.

It should be kept in mind that our sample may also be affected to some
extent by selection effects: for example, 11 of our Cepheids are
located in the vicinity of NGC 206, the giant star-forming region in
M31.  Additionally, M31 is observed at a rather high inclination
angle, which is not a typical situation for most galaxies searched for
Cepheids.  An opportunity to study the effects of blending with a
larger sample of $\sim100$ Cepheids, in a more face-on system, will
present itself in the next paper on Cepheids in M33 (Mochejska et
al. 2000). We will also discuss there the influence of blending on
the observed colors of Cepheids.

\section{Blending and the Light Curves}

The luminosity variations through the pulsation cycle of a Cepheid
seen in the optical are due primarily to changes in temperature.  The
monochromatic surface brightness changes with temperature and is
additionally quite dependent on wavelength. Of course, the Cepheid
also expands and contracts $-$ its area varies. This contributes to
the total light variation in a wavelength independent way.  The radius
variation in a typical Cepheid is out of phase by about 90$^{\circ}$
with respect to the surface brightness variation. Therefore a
comparison between color (proxy for temperature variation) and
magnitude (proxy for luminosity variation, due to {\em both} radius
and temperature change) would produce a loop in the CMD plane.

These loops, i.e. the tight correlation between color and luminosity,
have had a limited use in the past as part of the Cepheid PLC
calibration (Fernie 1964), or to transform away the
temperature-induced variation (Madore 1985; Moffett \& Barnes 1986).
The characterization of Cepheid companions had been traditionally done
by using the similar loops in the color-color plane of optical bands
(Stobie 1970). The color-color loop arises from the phase shift
introduced between the pair of color-magnitude relations when a
companion of different temperature (i.e. color) is added.  Blue or red
companions distort the color-color loops in very different and
quantifiable ways (Madore 1977; DeYoreo \& Karp 1979) and can be very
effective in deriving the amount of and color of contamination in a
Cepheids flux.

In order to understand the wavelength-dependent effect of a blue or
red companion on the light curves (and loops) of a Cepheid, one should
bear in mind that the contaminant flux is most prominent during the
minimum in the Cepheid lightcurve, and gradually becomes insignificant
as the Cepheid brightens up during its cycle.  The addition of a {\em
red companion} to the flux of a Cepheid has the following effects on
its optical light curves: (1) the light curve exhibits a flatter
minimum (due to the added flux); (2) the color curve has a deeper
minimum (due to the added red flux); and (3) the asymmetry in the
color curve decreases. The addition of a {\em blue companion} to the
flux of a Cepheid has the following effects on its optical light
curves: (1) the light curve exhibits a flatter minimum (as above); (2)
the color curve has a flatter minimum (due to the added blue flux);
and (3) the asymmetry in the color curve increases.

\begin{figure}[t]
\plotfiddle{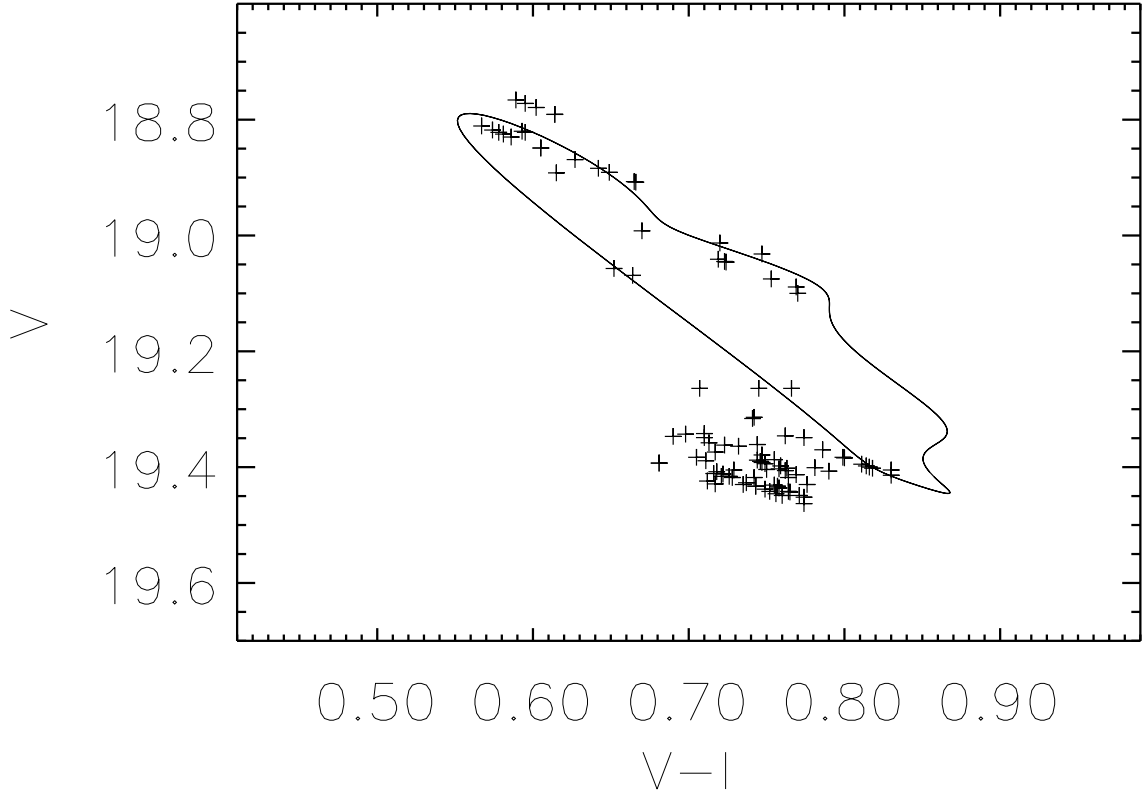}{10.3cm}{0}{100}{100}{-320}{-390}
\caption{The color-magnitude loop of the M31 Cepheid D31B V7184 from
our DIRECT observations (crosses), projecting a variation in
luminosity which is tightly correlated with the temperature
variation. For a single Cepheid the loop is traversed in a clockwise
direction, as shown by the model (solid).  The DIRECT observations of
V7184 include the constant flux of two unresolved blue B-stars $-$ the
effect on the loop is dramatic despite gaps in the phase coverage of
the light curves. Notice that when the Cepheid is bright (upper left
part) it is less affected by the blending and follows its loop as
expected.}
\label{fig:blend}
\end{figure}

A typical example is the Cepheid V7184 D31B (Paper I), which has blue
blends (Table 2) and whose CM loop is shown in Fig.~\ref{fig:blend}.
The effect on its loop is predictable given the above discussion: as
the Cepheid fades and becomes naturally redder, the added constant
blue flux of the blends becomes more prominent and diverts the lower
part of the loop to the blue (left in Fig.~\ref{fig:blend}). This
diversion could result eventually in a complete reversion of the
naturally clockwise trajectory of the loop. If the data is incomplete
(there are two gaps in the light curves of V7184), the blue blending
is still detectable by the steeper slope of the loop.

These changes on the color-magnitude plane, together with a fit of the
mean colors of the Cepheid to extinction laws, can be a powerful means
to find and characterize blended Cepheids in a sample like our DIRECT
project one. The necessary requirement is $-$ good, well-sampled light
curves in at least two bands. Therefore we were able to find blended
Cepheids like V7184 in M31B and exclude them from our analysis without
the benefit of {\em HST} images, but just using our light curves
(Paper I). However, the technique is limited to the cases with strongest
blending -- a nonlinear fitting routine by Rebel (1998) applied to the
M31B data still did not detect 2/3 of the blended Cepheids we found in
this paper. In other words, the naive technique of visual inspection
for "flat-bottom" light curves is completely inadequate for detecting
10\%-20\% Cepheid blending, which is dominating our findings in M31.

\section{Blend Analysis by the HST Key Project in the Galaxy NGC 2541}

The influence of blending has been largely neglected in many recent
galaxy distance determinations based on Cepheids. Few attempts have
been made to deal with this problem.  On one of the galaxies observed
by the HST Key Project on the Extragalactic Distance Scale, NGC 2541
(Ferrarese et al.~1998), several criteria were used to reject Cepheids
which may suffer from crowding. We have applied them to our sample of
22 Cepheids to estimate their effectiveness in detecting blending.
As we discussed already in \S 1, blending and crowding are different
phenomena, and judging from \S 5 they do not seem much correlated
in M31 either, but it would be helpful if these criteria could detect
blending.

\begin{figure}[t]
\plotfiddle{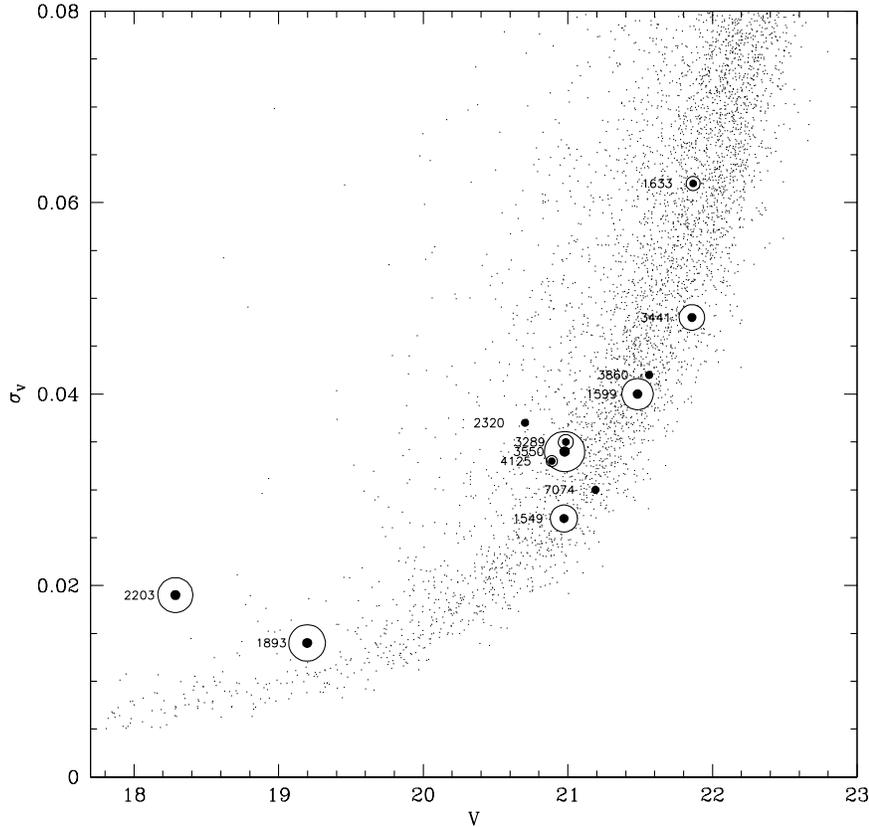}{10cm}{0}{60}{60}{-195}{-97}
\caption{The ALLSTAR photometric error $\sigma_V$ plotted as a
function of the ALLSTAR $V$ magnitude. The field F Cepheids are
denoted by solid dots surrounded by circles with their size
proportional to $S_V$ (no outside circle means $S_V=0$). The $V$
magnitudes and the errors are taken from the field F template image
(FWHM$\sim 1\arcsec$), as are the stars plotted in the background.}
\label{fig:als2}
\end{figure}

One of the proposed tests (Ferrarese et al.~1998) involves identifying
Cepheids that have companions not resolved by profile photometry
programs by looking at their photometric errors. The underlying idea
is that the photometric error would be higher than in case of an
isolated star of the same magnitude due to a poorer PSF fit resulting
from the unresolved companion.  We have constructed diagrams showing
the ALLSTAR photometric error $\sigma_V$ as a function of the ALLSTAR
$V$ magnitude, similar to the one presented in Fig.~8 of Ferrarese et
al.~(1998). The diagram for field F is shown in Figure \ref{fig:als2}.
The Cepheids are plotted as solid dots surrounded by circles with
their size proportional to the amount of blending $S_V$ (no outside
circle means $S_V=0$). The $V$ magnitudes and the errors are taken
from the field F template image (FWHM$\sim 1\arcsec$), as are the
stars plotted in the background.

Upon examining Figure \ref{fig:als2} it is apparent that the
photometric errors $\sigma_V$ do not show a correlation with the
amount of light coming from companions $S_V$. One such case is V2320
D31F which is an unblended star with an error almost twice as large
than most stars of similar magnitude. Another example is V3289 D31F
with $S_V=0.11$, which has a photometric error larger than two other,
more blended Cepheids of similar $V$ magnitude: V1549 D31F with
$S_V=0.28$ and V3550 D31F with $S_V=0.47$. Most of the Cepheids lie
within the most densely populated areas of the diagram, with V2320
D31F and V2203 D31F being the most clear outliers. A similar
photometric error distribution, uncorrelated with $S_V$, is also seen
in fields B and C.  In conclusion, the presence of close companions
did not manifest itself in the form of a poorer PSF fit and, hence, a
larger photometric error in the studied sample. Another fact which
should be kept in mind is that our ground-based PSF is much better
sampled than the {\em HST} PSF (comparing to the respective FWHMs),
thus it should be more sensitive to any deformations resulting from
the superposition of two or more stars.

Another proposed method of rejecting blended Cepheids is to remove all
stars having companions contributing more than $50\%$ of the total
light within a radius of two pixels. Taking into account the fact that
the two pixels on the {\em HST} WF chips correspond to twice the FWHM,
we have conducted a similar test on our ground-based $V$ template data
for the Cepheids. Only one of our Cepheids, V5646 D31B with $S_V=0.05$
would be rejected from our sample on the basis of this criterion. This
test also failed to identify the blended Cepheids in our sample.

A third proposed test for blending is to check if the Cepheid lies on
the instability strip in the CMD.  The usual color range spanned by
the final sample of Cepheids discovered with {\em HST} is about one
magnitude, for example $0.4\leq V-I\leq 1.4$ in NGC 2541 (Ferrarese et
al.~1998), $0.4\leq V-I\leq 1.5$ in NGC 4639 (Saha et al.~1997),
$0.6\leq V-I\leq 1.5$ in NGC 4535 (Macri et al.~1999). The examination
of a ground-based CMD for the 12 field F Cepheids yields a $V-I$ color
span of also about one magnitude ($0.6\leq V-I\leq 1.6$). No clear
correlation of the color with the $S_V$ of the Cepheid is seen in the
diagram, as unblended Cepheids also exhibit a large color scatter,
from $V-I=0.73$ for V2320 D31F to 1.59 for V3860 D31F, most likely due
to differential reddening.

In the case of our sample of 22 Cepheids, the tests were not
successful in rejecting the blends.  These tests may be helpful in
some cases of blending, when the two stars are almost resolved or when
the companion has a larger flux contribution than the maximum value of
$S_V=0.47$ present in our sample. But a contamination of $S_V=0.47$
has a large impact on the photometry of the Cepheid, which actually
would be dimmer by almost one third from the measured brightness.

\begin{figure}[t]
\plotfiddle{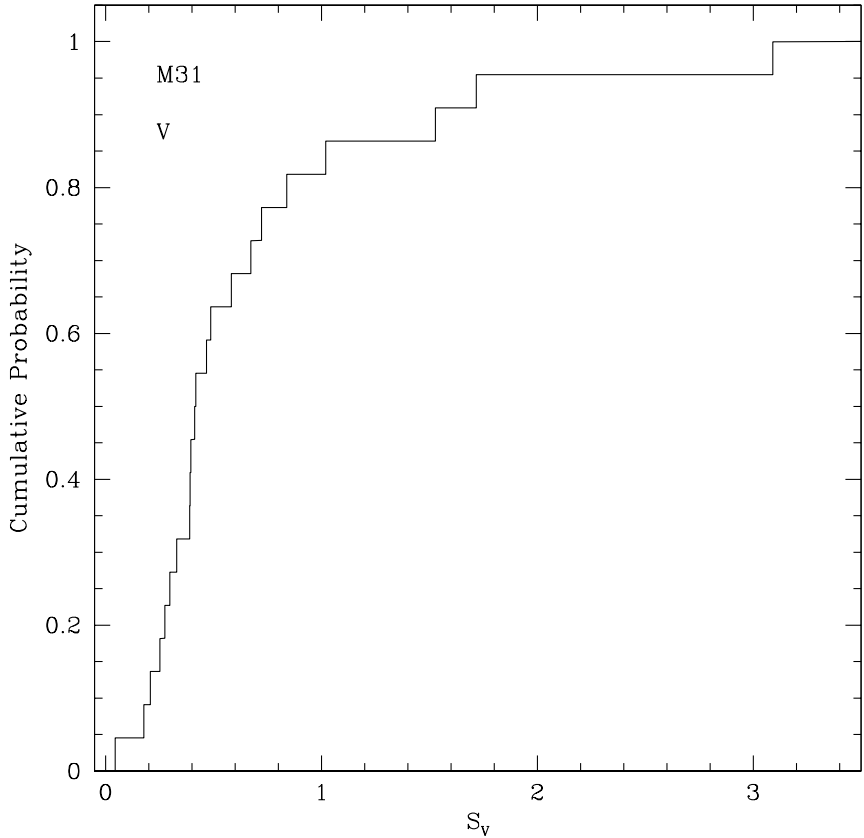}{10cm}{0}{60}{60}{-195}{-97}
\caption{The cumulative probability distribution of $S_V$ within a 
diameter of $3.2\arcsec$ of M31 Cepheids with {\em HST} data,
similar to that of one WF pixel ($0.1\arcsec$) at 25 Mpc.}
\label{fig:ks}
\end{figure}

\section{Conclusions}

For our sample of 22 Cepheids with both ground-based and {\em HST}
data, we find that the mean $V$-band flux contribution from companions
unresolved in the ground-based images, $\langle S_V \rangle$, is about
$19\%$ of the flux of the Cepheid, while the median $S_V$ is 12\%.
This shows that blending could potentially be a substantial source of
error in the Cepheid distance scale, as the distance derived from our
ground-based photometry for this admittedly small Cepheid sample would
be systematically underestimated by $\sim9\%$ (for the mean $S_V$) or
$\sim6\%$ (for the median $S_V$). This is to be compared to the
current Cepheid distance to M31 (which is subject to the LMC distance
uncertainty) of $R_{M31}= 770~\pm 25\; kpc$ (Freedman \& Madore 1990;
Kaluzny et al. 1998; Sasselov et al. 1998). The Cepheid photometry by
Freedman \& Madore (1990) was from the Canada-France-Hawaii telescope
(average seeing 1.0\arcsec, Freedman, Wilson, \& Madore 1991).  Our
findings of blending here require that these distance estimates to M31
be corrected upward by about 9\%.

Blending becomes even more severe when we consider galaxies at a
distance of $25\;Mpc$, i.e. at the edge of what can be currently
observed with {\em HST}. The $0.1\arcsec$ FWHM on the WF chips of the
WFPC2 camera would span a linear distance of $\sim12\;pc$ in such a
galaxy, which corresponds to $3.2\arcsec$ at the distance of M31. In
order to obtain an estimate as to the degree of contamination caused
by blending at a distance of $25\;Mpc$, we have summed the
contributions of all Cepheid companions within a diameter of
$3.2\arcsec$ in our {\em HST} data for M31.  The result is presented
in Figure \ref{fig:ks} in the form of a cumulative probability
distribution of $S_V$. The diagram shows that $50\%$ of our Cepheids
have $S_V > 0.4$, though some of the Cepheids with a very high degree
of contamination would probably elude detection.  This indicates that
blending will very likely introduce a significant contamination to
Cepheid photometry at such distances and resolution.

As the result of blending with other unresolved stars, the Cepheids
appear brighter than they really are when observed in distant galaxies
with {\em HST}.  As we compare them with mostly unblended LMC
Cepheids, this leads to systematically low distances to galaxies
observed with the {\em HST}, and therefore to systematically high
estimates of $H_0$.  The sign of the blending effect on the $H_0$ is
opposite to that caused by the lower LMC distance (e.g. Udalski 1998;
Stanek et al.~2000) and might be of comparable value, as discussed in
this paper. It should be stressed that blending is a factor which
contributes in only one direction, and therefore it will not average
out when a large sample of galaxies is considered.

One obvious solution to the problem of blending would be to obtain
data with better resolution. Such an opportunity will be available
after the launch of the Next Generation Space Telescope, scheduled for
2008. An alternative approach would be to determine the amount and
color of flux contamination for the Cepheids by the analysis of their
light curves and/or color-color loops. The requirement, however, is to
have good quality, well-sampled light curves in at least two bands,
which is not the case for {\em HST} Cepheids. However, even with our
good light curves it is difficult to detect Cepheids with blending of
$S_V<0.20$.

It must be stressed that the images taken with the {\em HST} WFPC2
camera, despite having a resolution $>10$ times better than our
ground-based images (FWHM $\sim0.4\;pc$ vs. $5\;pc$), will still leave
some blends unresolved, including physical companions (e.g. Evans
1992). Thus, the $V$-band flux contribution from Cepheid companions
derived in this paper sets only the lower limit on the true influence
of blending on the Cepheid photometry.

\acknowledgments{Bohdan Paczy\'nski, Andrzej Udalski and Thierry
Forveille have provided us with helpful comments on the manuscript. We
would like to thank Janusz Kaluzny for providing us with his database
management codes, Peter Stetson for the {\em HST} WFPC2 point-spread
functions and Doug Mink for assistance in automating image astrometry.
This work was partially based on observations with the NASA/ESA Hubble
Space Telescope, obtained from the data Archive at the Space Telescope
Science Institute, which is operated by the Association of
Universities for Research in Astronomy, Inc. under NASA contract
No. NAS5-26555. Support for this work was provided by NASA through
Grant AR-08354.02-97A from the Space Telescope Science Institute,
which is operated by the Association of Universities for Research in
Astronomy, Inc., under NASA contract NAS5-26555. BJM was supported by
the Polish KBN grants 2P03D00317 to Janusz Kaluzny and 2P03D01416 to
Grzegorz Pojma{\'n}ski.  DDS acknowledges support from the Alfred
P. Sloan Foundation. KZS was supported by the Harvard-Smithsonian
Center for Astrophysics Fellowship.}

\end{document}